\documentclass[journal]{IEEEtran}
\usepackage{cite}

\ifCLASSINFOpdf
  \usepackage[pdftex]{graphicx}
  \graphicspath{{figures}}
\else
\fi
\usepackage{amsmath}
\usepackage{amsfonts}
\usepackage{siunitx}
\usepackage{bbm}
\usepackage{pifont}
\newcommand{\cmark}{\ding{51}}  
\newcommand{\xmark}{\ding{55}}  
\usepackage{booktabs}

\ifCLASSOPTIONcompsoc
 \usepackage[caption=false,font=normalsize,labelfont=sf,textfont=sf]{subfig}
\else
 \usepackage[caption=false,font=footnotesize]{subfig}
\fi
\begin{document}
%
\title{Predicting Tweet Posting Behavior on Citizen Security: A Hawkes Point Process Analysis}

\vspace{0.5em}
\large \textit{This work has been submitted to IEEE Transactions on Network Science and Engineering for possible publication. Copyright may be transferred without notice, after which this version may no longer be accessible.}

%
%
%

\author{Cristian~Pulido
        and~Francisco~Gómez%
\thanks{Cristian Pulido is with the Departamento de Ingeniería de Sistemas e Industrial, Universidad Nacional de Colombia, Bogotá, Colombia (e-mail: cpulido@unal.edu.co).}%
\thanks{Francisco Gómez is with the Departamento de Matemáticas, Universidad Nacional de Colombia, Bogotá, Colombia (e-mail: fagomezj@unal.edu.co).}%
}

%
%

\maketitle

\begin{abstract}
The Perception of Security (PoS) refers to individuals' subjective sense of safety in specific contexts. While surveys have traditionally captured these perceptions, they are limited in frequency, timeliness, and predictive capability. Recent advances suggest that social media data offer a complementary means to monitor PoS more dynamically. However, the challenge of anticipating PoS trends in real time remains underexplored. This paper presents a novel predictive framework for short-term forecasting of PoS-related activity using Twitter data. The proposed model integrates external covariates, such as social protests and sports events, into a Hawkes point process, enabling the estimation of future tweet volumes related to PoS. Natural language processing and machine learning techniques are employed to quantify PoS from text and enhance model input. Empirical results on data from Bogotá, Colombia, show that the model achieves competitive predictive accuracy while offering interpretability through covariate weighting. This approach provides actionable insights into the temporal drivers of public sentiment, supporting proactive security planning and resource allocation. Overall, this work contributes to the integration of artificial intelligence techniques in engineering applications aimed at social impact.
\end{abstract}


\section{Introduction}
\IEEEPARstart{T}{he} Perception of Security (PoS) refers to the subjective feeling of safety or insecurity that individuals experience in specific places or situations~\cite{Skogan1981}. Negative PoS can lead to stigmatization of communities, mental health issues, distrust in institutions, and economic losses~\cite{risk}. These perceptions are shaped by both personal experience and contextual events such as crime, social unrest, and mass gatherings~\cite{skogan1993,konstantaki2010residents,eraslan2023spectators}.

\noindent Traditionally, PoS has been measured via periodic surveys~\cite{boholm1998comparative,ladenburg2021watching}. While valuable, this approach lacks the temporal resolution and scalability required for real-time monitoring or short-term forecasting. In contrast, social media platforms such as Twitter form large-scale, dynamic information networks, where users interact through posting, retweeting, and responding, enabling the emergence and propagation of collective perceptions in near real time~\cite{chaparro2021quantifying,curiel,greco2021security}.

\noindent From a network science perspective, these platforms can be viewed as evolving networks where nodes (users) and edges (interactions) produce observable temporal patterns~\cite{tumasjan2010predicting}. The temporal dynamics of content diffusion and user influence reflect underlying network structure, even if not explicitly reconstructed~\cite{twitterimpact}. In this context, point process models such as the Hawkes process provide a principled way to capture self-excitation, cascades, and the influence of contextual covariates on event generation~\cite{santitissadeekorn2025influence}.

\noindent In this work, we hypothesize that short-term changes in PoS can be anticipated by combining online sentiment signals with contextual factors such as protests and sports events. These elements shape the tempo and topology of discourse, affecting both the rate and spread of PoS-related content. We present an interpretable, data-driven framework based on a covariate-enhanced Hawkes process to predict the volume of PoS-related tweets in Bogotá, Colombia. Results show that the model achieves competitive accuracy compared to baselines while offering explanatory insights into the drivers of public sentiment.

\section{Related Work}
\label{sec:related_work}

This section reviews prior contributions on: (1) social media-based analysis of security perception, (2) predictive modeling of event dynamics, and (3) integration of contextual covariates in forecasting. Table~\ref{tab:related_works} summarizes whether each work addresses Perception of Security (PoS), includes predictive components, and incorporates contextual factors.

\subsection{Social Media and Security Perception}

Early work by Prieto and Bishop~\cite{prieto2017modelling} modeled fear propagation as a dynamic process in urban spaces, while Curiel et al.~\cite{curiel} examined reactions to crime on social media. Chaparro et al.~\cite{chaparro2021quantifying} introduced a tweet-level classification pipeline to quantify PoS in Bogotá. These studies focus on retrospective quantification of sentiment but do not address forecasting.

\noindent More recently, Messa et al.~\cite{messa2025modeling} analyzed spatial determinants of perceived safety for women in Milan using regression over survey data. While insightful, their work is static and lacks temporal modeling. Su et al.~\cite{Su2025DiffusionNetworks} compared diffusion networks of public and entertainment events, showing how certain topics sustain attention longer. Although not specific to PoS, this highlights the networked nature of attention dynamics, reinforcing the need for models that account for cascades and influence patterns.

\subsection{Predictive Modeling and Network Dynamics}

Social media has supported predictions in areas like elections~\cite{tumasjan2010predicting}, health~\cite{marques2017dengue}, and finance~\cite{balfagih2019evaluating}. However, most methods rely on regression and neglect the event-driven, temporally structured nature of interactions. In contrast, Hawkes processes offer a network-aware formulation where each event can trigger future ones. Zhao et al.~\cite{zhao2015seismic} and Kobayashi and Lambiotte~\cite{kobayashi2016tideh} modeled retweet cascades as self-exciting processes. Rizoiu et al.~\cite{rizoiu2017tutorial} provided a general overview, and Santitissadeekorn et al.~\cite{santitissadeekorn2025influence} applied multidimensional Hawkes processes to infer latent networks from count data. Yet, these approaches often lack semantic interpretability and do not incorporate external covariates.

\subsection{Contextual Covariates in Forecasting}

The influence of exogenous factors on perceived security is well established~\cite{skogan1993,shibata2024effect}. Reinhart~\cite{reinhart2016point} integrated spatiotemporal covariates into point process models for crime prediction. Messa et al.~\cite{messa2025modeling} also included urban features such as lighting and land use. However, these models do not address the dynamics of user-generated content or the self-exciting nature of online discourse.

\subsection{Comparison with Prior Work}

As shown in Table~\ref{tab:related_works}, while prior research has addressed PoS, predictive modeling, or covariate integration individually, our work is the first to unify all three using an interpretable, event-driven approach based on Hawkes processes. It models the dynamics of PoS-related discourse on social media, influenced by both endogenous retweet behavior and exogenous urban events.

\begin{table}[htbp]
\centering
\scriptsize
\caption{Comparison of Related Works Addressing PoS and Predictive Modeling}
\label{tab:related_works}
\begin{tabular}{lccc}
\toprule
\textbf{Reference} & \textbf{PoS} & \textbf{Predictive} & \textbf{Covariates} \\
\midrule
Prieto \& Bishop (2017)~\cite{prieto2017modelling} & \cmark & \xmark & \cmark \\
Greco \& Polli (2021)~\cite{greco2021security} & \cmark & \xmark & \cmark \\
Chaparro (2021)~\cite{chaparro2021quantifying} & \cmark & \xmark & \xmark \\
Messa et al. (2025)~\cite{messa2025modeling} & \cmark & \xmark & \cmark \\
Su et al. (2025)~\cite{Su2025DiffusionNetworks} & \xmark & \xmark & \cmark \\
Rizoiu (2017)~\cite{rizoiu2017tutorial} & \xmark & \cmark & \xmark \\
Kobayashi \& Lambiotte (2016)~\cite{kobayashi2016tideh} & \xmark & \cmark & \xmark \\
Santitissadeekorn et al. (2025)~\cite{santitissadeekorn2025influence} & \xmark & \cmark & \xmark \\
\textbf{This work} & \cmark & \cmark & \cmark \\
\bottomrule
\end{tabular}
\end{table}

\section{Materials and Methods}

\begin{figure*}
\centering
\includegraphics[width=0.55\textwidth]{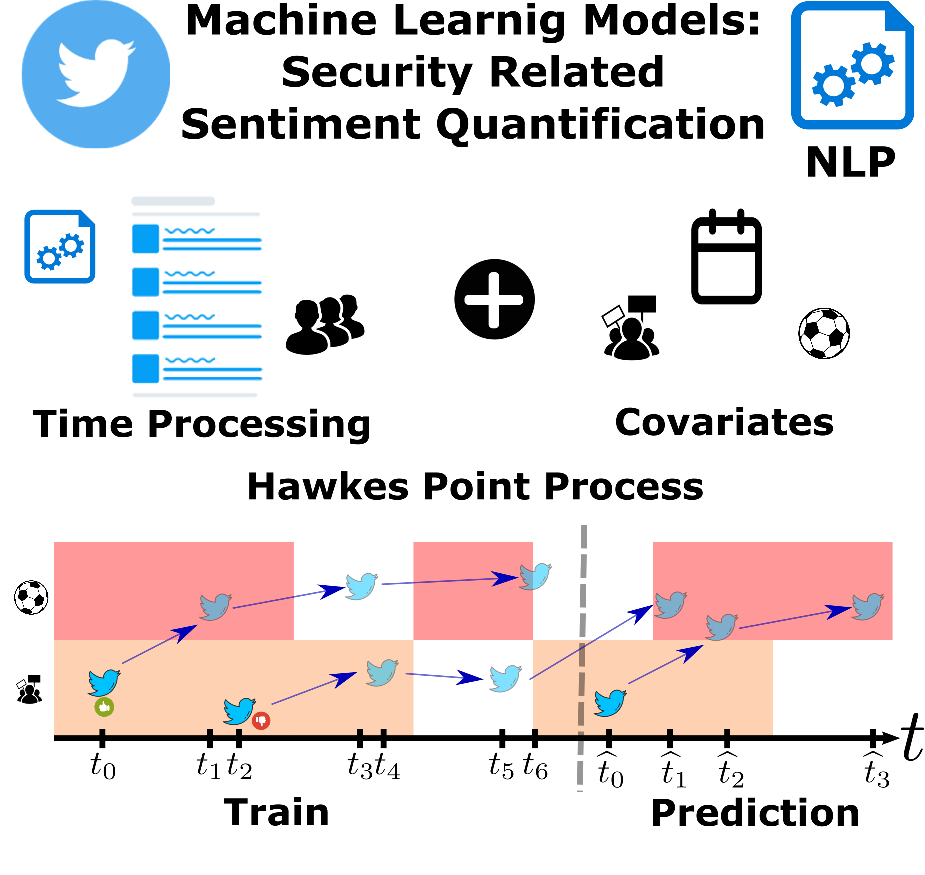}
\caption{Schematic representation of the proposed predictive model. The system integrates tweet data, classification and quantification of PoS-related content, and exogenous temporal covariates into a prediction-enabled Hawkes process.}
\label{fig:generalmodel}
\end{figure*}

Figure~\ref{fig:generalmodel} outlines the architecture of the proposed model. This work builds on the strategy introduced in~\cite{chaparro2021quantifying}, where social media content was classified to estimate the Perception of Security (PoS). It also expands upon an earlier version presented at CIARP 2021~\cite{Pulido2021,tavares2022progress}, incorporating richer features and predictive capabilities.

\noindent The model uses as input: (i) tweets labeled as PoS-related via supervised text classification, (ii) metadata such as timestamps, retweet structure, and user follower counts, and (iii) exogenous temporal covariates such as protests and football matches. This heterogeneous information is embedded into a Hawkes point process, which models both the self-exciting nature of information cascades and the external modulation induced by contextual events.

\noindent Parameters of the Hawkes process are fitted via maximum likelihood on historical data. Once trained, the model can simulate future activity and anticipate the volume of PoS-related discourse over short time horizons. This supports proactive monitoring of fluctuations in perceived security, driven by both endogenous (online propagation) and exogenous (real-world events) influences.

\subsection{Social Network Data}

Twitter data geolocated in Bogotá, Colombia, were collected using Twitter’s Streaming API between March 18, 2019, and April 28, 2020, yielding a dataset of 1,701,668 tweets. This period corresponds to the timeframe defined within the scope of the city-funded project for predictive analytics in security and coexistence (BPIN: 2016000100036), which aimed to explore computational approaches for anticipating public sentiment related to safety.

\noindent As described in~\cite{chaparro2021quantifying}, tweets were filtered based on a predefined lexicon of security-related terms. A manually labeled subset was annotated with sentiment scores on a five-level scale (1: very negative to 5: very positive), and a supervised classification model (multinomial Naive Bayes with Bag-of-Words representation) was trained to automatically score the remaining posts.

\noindent In addition to sentiment labels, each tweet was enriched with metadata including its timestamp and the number of followers of the posting account. These features served as inputs to the predictive model.

\noindent To capture external factors affecting PoS expression, two exogenous covariates were incorporated: (i) dates of protests and public demonstrations, known to influence perceived safety by altering public order and institutional presence~\cite{Skogan1981,skogan1993}; and (ii) dates of soccer matches involving Bogotá-based teams (Team A and Team B), which have been linked to localized incidents of crime and public disturbances~\cite{ayres2012bars,ristea2020spatial,news_crime_football}.

\subsection{Proposed Model}

In the digital age, individuals increasingly use social media to express opinions related to their Perception of Security (PoS)~\cite{curiel}. Among these platforms, Twitter (now known as X) plays a central role in the diffusion of public sentiment through original content and retweets~\cite{twitterimpact,microblog}. This diffusion dynamic motivates the modeling of tweet activity as a time-dependent process that captures both direct user participation and the reverberation of shared content.

\noindent The objective of the proposed model is to forecast the hourly volume of security-related tweets, comprising both original posts and retweets, within short-term horizons. The forecasting task is framed as a discrete-time prediction problem, where the future tweet volume depends on historical activity and contextual covariates such as protests or sporting events~\cite{zhao2015seismic,kobayashi2016tideh}.

\noindent To capture these dynamics, a Hawkes point process with covariates is employed~\cite{rizoiu2017tutorial}. This self-exciting temporal model accounts for both exogenous influences and the endogenous propagation of posts. The event intensity function $\lambda(t)$ at time $t$ is defined as:

\begin{equation*}
\lambda(t) = \mu(t) + \sum_{i: t_i < t} g(t - t_i),
\end{equation*}

\noindent
where $\mu(t)$ is the baseline rate of original tweets, modulated by covariates, and $g(\cdot)$ is a kernel function representing the excitation effect from past events. This formulation allows the model to represent how external events shift tweet generation rates, while also capturing bursty behavior driven by user interactions and retweets~\cite{reinhart2016point}.

\subsubsection{Original Tweet Posting}

The background intensity term $\mu(t)$ models the rate of original tweets related to the Perception of Security (PoS)~\cite{curiel}. It is defined as an exponential function over a set of temporal and contextual covariates:

\begin{equation*}
\mu(t) = \exp\left( \boldsymbol{\beta}^\top \mathbf{C}(t) \right),
\end{equation*}

\noindent
where $\boldsymbol{\beta}$ is a vector of learned coefficients, and $\mathbf{C}(t)$ is a covariate vector encoding features relevant to posting behavior~\cite{reinhart2016point}. The covariates include:

\begin{itemize}
    \item \textbf{Day of the week} (integer from 0 to 6),
    \item \textbf{Time of day}, captured via binary indicators for 00:00–12:00 and 12:00–24:00,
    \item \textbf{Event indicators}: binary variables marking the occurrence of (i) soccer matches involving local teams, and (ii) civic demonstrations such as protests or rallies.
\end{itemize}

\noindent These features capture both regular diurnal patterns in user activity~\cite{microblog} and abrupt variations driven by exogenous events known to impact public sentiment and PoS expression~\cite{skogan1993}.

\subsubsection{Retweet Posting}

Unlike original tweets, retweets follow distinct propagation dynamics~\cite{microblog}. For instance, posts by influential users are more likely to be rapidly and widely replicated~\cite{twitterimpact}. Moreover, the influence of a tweet diminishes over time, depending on its relevance and the reach of users who amplify it.

\noindent In the context of PoS-related content, negatively charged messages expressing insecurity or fear, tend to propagate more rapidly, reinforcing negative perceptions in the population~\cite{prieto2017modelling}. Conversely, positive or reassuring messages often have limited impact. To capture these dynamics, the contribution of a past event $i$ to the overall event intensity at time $t$ is modeled as:

\begin{equation*}
g(t - t_i) = p_i(t) \sum_{j \in RT(i)} d_j\, \psi(t - t_j),
\end{equation*}

\noindent where $RT(i)$ denotes the set of tweets (including the original) associated with event $i$, $d_j$ is the follower count of user $j$ who posted at time $t_j$, and $\psi(s)$ is a memory kernel function that modulates the temporal decay in influence. In this work, the functional form of $\psi(s)$ is adopted directly from prior empirical studies on tweet diffusion dynamics~\cite{kobayashi2016tideh,rizoiu2017tutorial}, where it was shown to capture both the initial burst and the long-tail behavior of retweet activity. Specifically, the kernel is defined as:

\begin{equation*}
\psi(s) =
\begin{cases}
0 & \text{if } s < 0, \\
6.49 \times 10^{-4} & \text{if } 0 \leq s \leq 300, \\
5.44 \times 10^{-7} \cdot s^{-1.242} & \text{if } s > 300,
\end{cases}
\end{equation*}

\noindent capturing both short-term surges and long-tail dynamics of retweet activity. The function $p_i(t)$ modulates the influence of the original tweet over time, accounting for content sentiment and temporal decay:

\begin{equation*}
\begin{split}
p_i(t) =\; & p_0^i \left[ 1 - (S_i r_0) \sin\left( \frac{2\pi}{T_m}(t + \phi_0) \right) \right] \\
& \times \exp\left( -\frac{t - t_0}{\tau_m} \right),
\end{split}
\end{equation*}

\noindent where $S_i$ represents the sentiment of the tweet (e.g., negative vs. positive), and the sinusoidal and exponential terms capture the influence of circadian cycles and information aging, respectively. The parameters $p_0^i$, $r_0$, $T_m$, $\phi_0$, and $\tau_m$ are learned from historical data.

\noindent
This function represents an exponentially decaying oscillatory pattern, modeled as the product of sinusoidal and exponential components. In this expression:
\begin{itemize}
    \item $S_i$ is the PoS value associated with the original tweet,
    \item $p_0^i$ is the baseline influence intensity of tweet $i$,
    \item $t_0$ denotes the timestamp of the original tweet,
    \item $T_m$ is the period of oscillation, fixed to one day~\cite{kobayashi2016tideh},
    \item $r_0$ is the relative amplitude of the oscillation,
    \item $\phi_0$ is the phase shift,
    \item $\tau_m$ is the characteristic decay time~\cite{kobayashi2016tideh}.
\end{itemize}

\noindent
The set $P_0 = \{p_0^i \mid i \in OT\}$ comprises the baseline intensities for all original tweets, where $OT$ is the index set of original tweets. The parameters to be estimated during model training include $P_0$, $r_0$, $\phi_0$, $\tau_m$, and $\boldsymbol{\beta}$.

\begin{figure}[htbp]
  \centering
  \subfloat[$\psi$ function.]{%
    \includegraphics[width=0.9\linewidth]{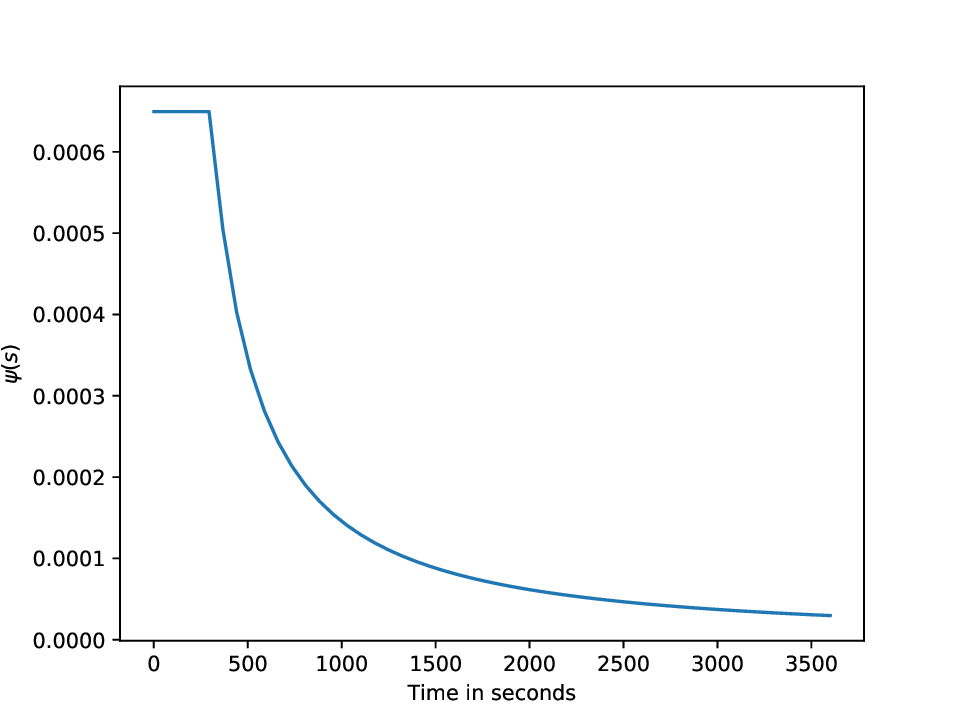}%
    \label{fig:psi}
  }\\ 
  \subfloat[Influence function $p(t)$ for PoS values $S \in \{1,3,5\}$.]{%
    \includegraphics[width=0.9\linewidth]{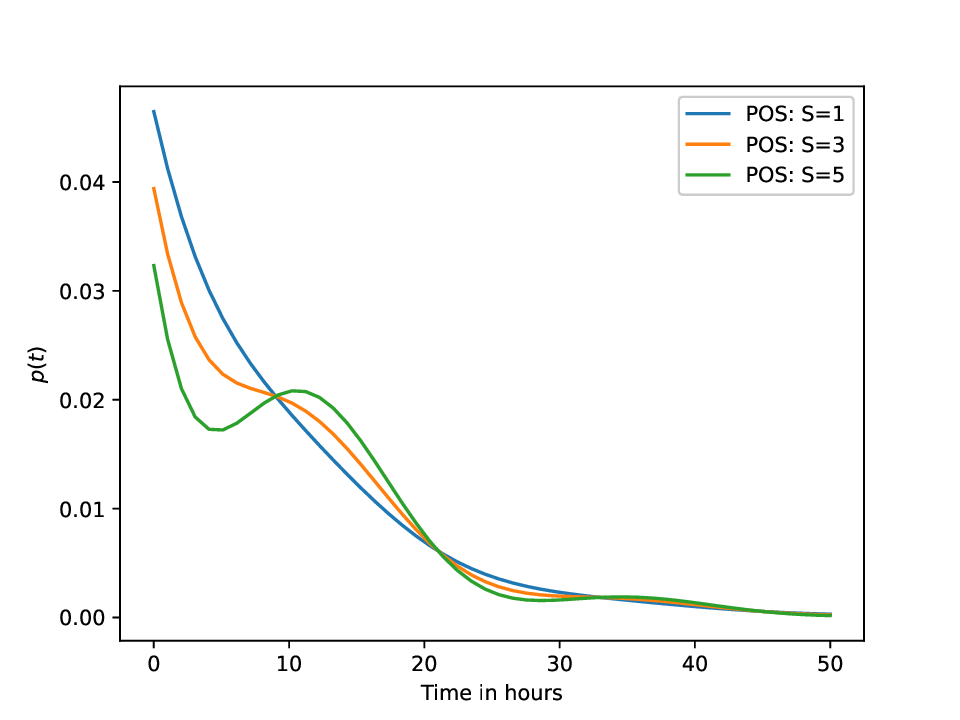}%
    \label{fig:influence}
  }
  \caption{Behavior of the kernel function $\psi$ and the influence function $p(t)$. The function $\psi$ starts as a constant and then decays rapidly, modeling retweet likelihood over time. The influence function $p(t)$ exhibits a decaying oscillatory pattern modulated by the PoS value $S$.}
  \label{fig:examples_functions}
\end{figure}

\noindent Figure~\ref{fig:examples_functions} presents the temporal behavior of the functions $\psi$ and $p(t)$. In panel~\ref{fig:psi}, $\psi(s)$ is shown to remain nearly constant immediately after an original tweet is posted and then decay rapidly, modeling the diminishing likelihood of retweeting over time. Panel~\ref{fig:influence} depicts how different PoS values ($S \in \{1,3,5\}$) affect the influence function $p(t)$. In these examples, the parameters were set as $p_0 = 0.05$, $r_0 = 0.1$, $\phi_0 = 0.3$, and $\tau_m = 10$. Tweets with more negative PoS values (e.g., $S=1$) initially exhibit stronger influence. Over time, the function follows a damped sinusoidal trajectory, reflecting the decaying impact of the original tweet.

\subsection{Estimating Model Parameters}

The model parameters, denoted by $\Theta = \{P_0, r_0, \phi_0, \tau_m, \boldsymbol{\beta}\}$, were estimated from historical data using the maximum likelihood approach proposed in~\cite{reinhart2016point}. The following subsections describe the estimation process in detail; additional derivations and mathematical formalizations are provided in Appendix~\ref{anexo:modelo_matematico}. To compute $\Theta$, the log-likelihood function $\ell(\Theta)$ over the training time interval $[t_a, t_b]$ was formulated as:

\begin{equation*}
\label{eq:loglikelihood}
\begin{split}
\ell(\Theta) =\; & \sum_{i \in OT} \boldsymbol{\beta}^\top \mathbf{C}(t_i) \\
& + \sum_{i \in OT} \sum_{j \in RT(i)} \ln\left( p_i(t_j) \, d_j \, \psi(t_j - t_i) \right) \\
& - \int_{t_a}^{t_b} \lambda(t)\, \mathrm{d}t.
\end{split}
\end{equation*}

\subsubsection{Estimating Background Parameters}

The parameter vector $\boldsymbol{\beta}$ influences only the first term of the log-likelihood expression and the first component of the integral term when substituting $\lambda(t)$. By computing the derivative of $\ell(\Theta)$ with respect to $\boldsymbol{\beta}$ and setting it to zero, we obtain:

\begin{equation}
\label{eq:beta_derivative}
\frac{\partial \ell}{\partial \boldsymbol{\beta}} = \sum_{i \in OT} \mathbf{C}(t_i) - \int_{t_a}^{t_b} \mathbf{C}(t) \exp\left( \boldsymbol{\beta}^\top \mathbf{C}(t) \right)\, \mathrm{d}t = 0.
\end{equation}

\noindent
Assuming that the covariate function $\mathbf{C}(t)$ is piecewise constant over a partition $\mathcal{P}$ of the interval $[t_a, t_b]$, we approximate the integral in~\eqref{eq:beta_derivative} as a weighted sum:

\begin{equation}
\label{eq:beta_discretized}
\sum_{i \in OT} \mathbf{C}(t_i) = \sum_{p_i \in \mathcal{P}} \mathbf{C}(\overline{p}_i)\, \exp\left( \boldsymbol{\beta}^\top \mathbf{C}(\overline{p}_i) \right) \, |p_i|.
\end{equation}

\noindent
Here, $|p_i|$ denotes the length of partition element $p_i$, and $\overline{p}_i$ is its midpoint, used to represent the covariate vector within that interval. Equation~\eqref{eq:beta_discretized} does not yield a closed-form solution for $\boldsymbol{\beta}$; therefore, numerical optimization techniques were applied to estimate its value. As noted in~\cite{reinhart2016point}, the likelihood function is convex in $\boldsymbol{\beta}$, which ensures stability and convergence of the optimization process.

\subsubsection{Estimating Influence Function Parameters}

To estimate the remaining parameters associated with the influence function, we adopt the assumption that a tweet's influence remains approximately constant within small time windows, following the approach in~\cite{kobayashi2016tideh}. Accordingly, we approximate $p_i(t)$ by a discretized version $\widehat{p}_i(t)$, computed over uniform intervals of approximately four hours. This approximation enables the use of a maximum likelihood-based estimator, from which the parameters $P_0$, $r_0$, $\phi_0$, and $\tau_m$ are obtained by minimizing an error function.

\noindent Let $i \in OT$ denote an original tweet posted at time $t^i_0$, and let $[t_{\mathrm{st}}, t_{\mathrm{end}}]$ be a time window in which tweet $i$ receives $R$ retweets. The discretized influence is defined as:

\begin{equation*}
\label{eq:phat}
\widehat{p}_i(t) = R \left( \sum_{j \in RT(i)} d_j \int_{t_{\mathrm{st}} - t_j}^{t_{\mathrm{end}} - t_j} \psi(t) \, \mathrm{d}t \right)^{-1}.
\end{equation*}

\noindent Let $T(i)$ be the partition of the interval $[t_0^i, t_b]$ into subintervals over which $\widehat{p}_i(t)$ is considered constant. At the midpoint $\overline{t}_j$ of each interval in $T(i)$, the difference between the estimated influence $\widehat{p}_i$ and the modeled influence $p_i(t)$ is evaluated. The total error function is defined as:

\begin{equation*}
\label{eq:error}
E(P_0, r_0, \phi_0, \tau_m) = \sum_{i \in OT} \sum_{\overline{t}_j \in T(i)} \left\| \widehat{p}_i(\overline{t}_j) - p_i(\overline{t}_j) \right\|.
\end{equation*}

\noindent
This function is minimized to estimate the optimal values of the parameters $P_0$, $r_0$, $\phi_0$, and $\tau_m$, which define the influence function $p_i(t)$. These estimates are then used in the final predictive model to compute the future posting intensity.

\subsection{Predicting Future Tweets}

The intensity function $\lambda(t)$ can be directly evaluated for any $t \in [t_a, t_b]$ using the observed data. However, for any future time $t > t_b$, the retweet times, sentiment polarity (PoS), and follower counts are unknown. To extend $\lambda(t)$ into the prediction window, we estimate the additional expected intensity generated by future original tweets and their anticipated retweets.

\noindent The procedure begins by generating samples of future background events using the thinning algorithm~\cite{rizoiu2017tutorial} applied to the function $\mu(t)$. This process yields a set of predicted original tweet times, denoted as $OTP$. For each synthetic tweet $i \in OTP$, we assign a PoS value, a baseline intensity $\overline{p}_0^i$, and a number of followers $d_i$, sampled from the empirical distributions observed in the training data.

\noindent These sampled values allow us to compute the expected additional influence generated by each predicted tweet. The total expected future intensity at time $t > t_b$ is given by:

\begin{equation*}
\label{eq:lambda_prediction}
\begin{split}
\widehat{\lambda}(t) =\; & \lambda(t) \\
& + \sum_{i \in OTP} \mathbb{E}[p_i(t)] \cdot \mathbb{E}[d_i] \int_{t_b}^{t} \psi(t - s)\, \mathrm{d}s.
\end{split}
\end{equation*}

\noindent
Once the intensity $\widehat{\lambda}(t)$ is obtained for a target interval $[t_b, t]$, the thinning algorithm is applied again to generate a realization of the predicted number of tweets related to security during that period.

\subsection{Experimental Settings}

To assess the predictive capacity of the proposed model, two baseline methods were implemented for comparison. The first was a non-homogeneous Poisson process (NHPP)~\cite{lawless1987regression}, which modeled the aggregated number of events (tweets and retweets combined) using the same covariates as explanatory variables. The second baseline extended this by decoupling the modeling of original tweets and retweets: a regression model was trained to predict the logarithm of the number of retweets per original tweet, based on its early activity, following the strategy in~\cite{kobayashi2016tideh}. The original tweet intensity was estimated using the background component $\mu(t)$ defined in this work, while retweet counts were extrapolated from their initial propagation behavior.

\noindent
To simulate realistic short-term forecasting scenarios, a time-based cross-validation scheme was employed. The complete dataset was partitioned into consecutive, non-overlapping 15-day blocks. At each iteration, the model was trained on a 30-day sliding window and evaluated on the subsequent 15-day block. This procedure preserved temporal ordering and avoided information leakage, thus mimicking a prospective prediction setup.

\noindent
Model performance was quantitatively evaluated using two complementary metrics: the Mean Absolute Error (MAE), which captures average deviations between predicted and observed tweet counts, and the Pearson correlation coefficient, which measures the strength of linear association~\cite{willmott2005advantages}. These metrics were computed over hourly prediction intervals.

\noindent
Beyond accuracy, the proposed model's interpretability was examined through an analysis of the estimated coefficients for temporal and contextual covariates. This allowed for identifying key drivers of PoS-related activity and understanding how specific factors, such as day of the week, time of day, or the occurrence of protests, modulate the dynamics of online security discourse.

\noindent
Although the data used span 2019–2020, they originate from a large-scale project funded to explore predictive analytics for urban safety (see Acknowledgment). The modeling framework, however, is agnostic to the specific time period and generalizable to other cities or more recent datasets, provided comparable covariates and labeled content are available.

\section{Results}

This section presents the empirical evaluation of the proposed model across multiple temporal windows and event contexts. The analysis focuses on three core aspects: (i) the predictive alignment between model estimates and observed tweet volumes, (ii) the learned influence of contextual covariates such as protests and sports events, and (iii) comparative performance against established baseline models. In addition to accuracy metrics, we also examine the interpretability afforded by the model’s structure, particularly through the analysis of fitted covariate weights that explain fluctuations in Perception of Security (PoS)-related discourse over time.

\subsection{Covariates' Effects}

\begin{figure}[t]
\centering
\includegraphics[width=0.48\textwidth]{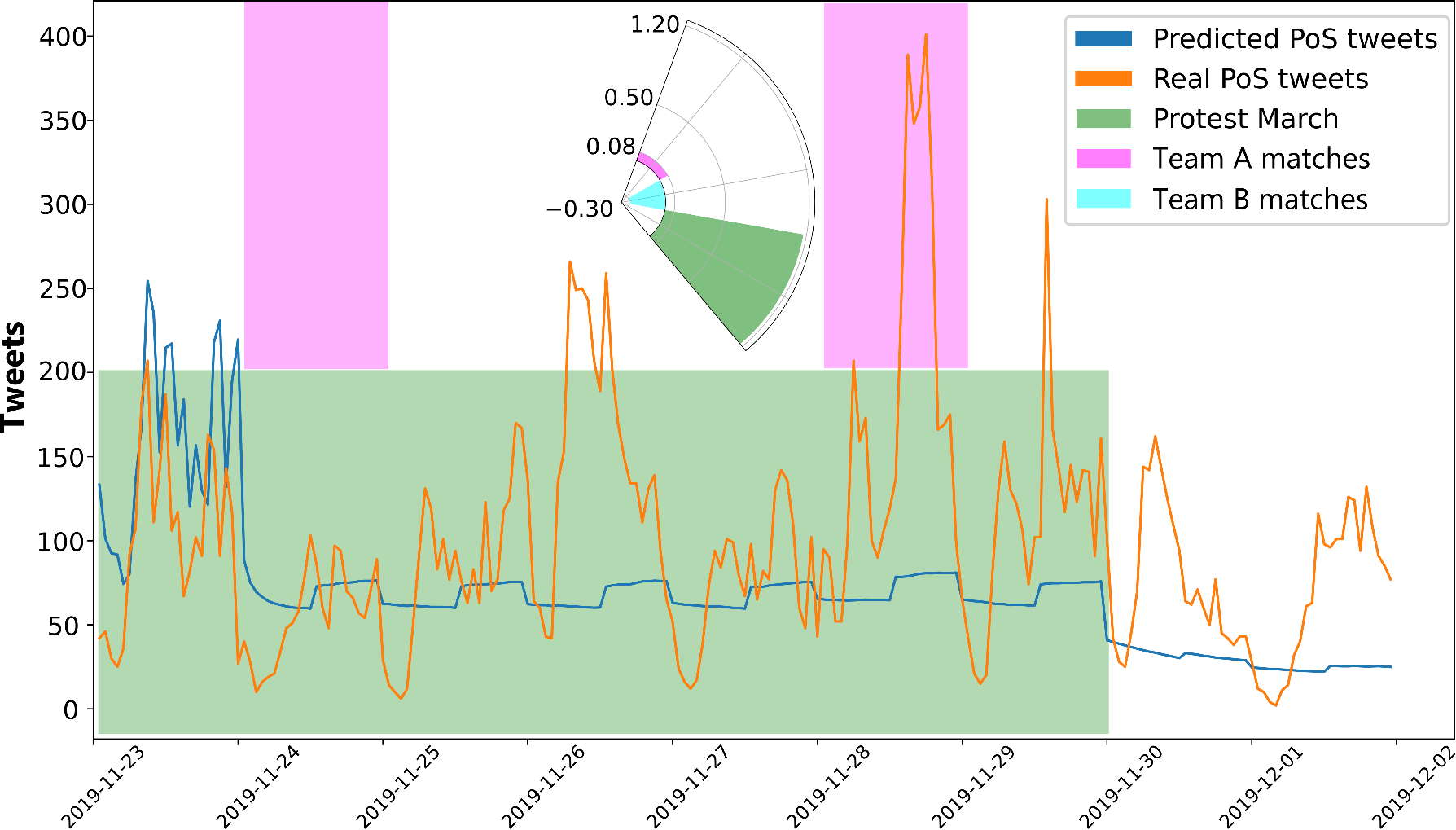}
\caption{Predicted vs. observed number of security-related tweets from November 23 to December 1, 2019. The model was trained on six months of data. The blue line represents predictions; the orange line denotes actual tweet counts. Colored bands highlight days with key covariates: green (protests), fuchsia (Team A matches), and aquamarine (Team B matches). Radial bars depict covariate weight contributions.}
\label{fig:2019}
\end{figure}

Figure~\ref{fig:2019} illustrates the prediction of tweets related to Perception of Security (PoS) over a nine-day horizon, using the proposed model. The training set comprised six months of data, from May 23 to November 23, 2019, starting at midnight. This period was selected due to the presence of multiple social protests and local soccer matches in Bogotá, which were hypothesized to affect PoS-related tweet volume. In the figure, the $x$-axis is segmented by 24-hour intervals for each day, providing hourly resolution. The blue curve represents the model’s predicted number of PoS-related tweets (including both original posts and retweets), while the orange curve shows the actual observed counts. 

\noindent Colored bands are overlaid to indicate the occurrence of key covariates: Green: days with social protests, Fuchsia: Team A soccer matches,  Aquamarine: Team B soccer matches. These visual cues help contextualize changes in tweet volume with respect to external events. Additionally, radial bars are included to reflect the relative weights assigned to each covariate by the model, offering insight into their contribution to predicted activity.

\noindent Furthermore, Figure~\ref{fig:2019} includes a radial bar plot that displays the fitted weights assigned to each covariate. These weights quantify the relative contribution of each factor to the predicted tweet intensity. The radial plot provides a visual summary of the covariates’ influence within the model. As shown in Figure~\ref{fig:2019}, the proposed model effectively captures the oscillatory patterns observed in the actual tweet data related to PoS. While the timing and general shape of these fluctuations are well reproduced, the model tends to underestimate the amplitude of the peaks. This indicates a limitation in modeling extreme surges in activity.

\noindent A notable example occurs on November 28, 2019, where the simultaneous presence of two covariates influences the prediction, citizen protests and a soccer match. The model attributes a significant increase in tweet activity to the protest event, highlighting its relevance in shaping public discourse on security. In contrast, the soccer match appears to have a marginal effect on the predicted tweet volume. This suggests that, during this specific period, social protests were more impactful than local sports events in driving online discussions about perceived security.

\begin{figure}[t]
\centering
\includegraphics[width=0.48\textwidth]{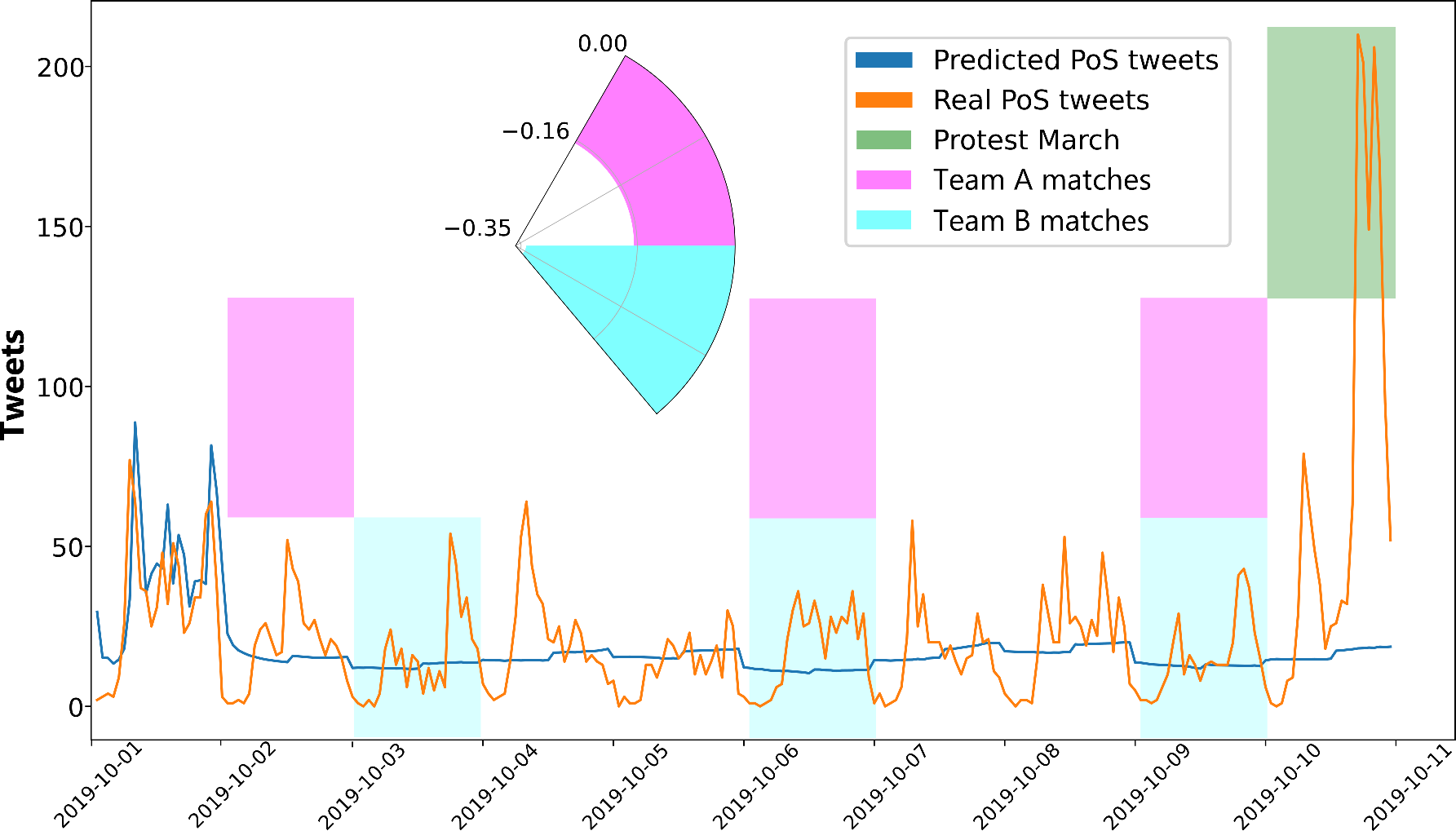}
\caption{Predicted vs. observed tweet volume from October 1 to 10, 2019, using six months of training data. The orange line denotes the actual number of security-related posts; the blue line shows the model’s predictions. Colored bands mark key covariates: green (protests), fuchsia (Team A matches), and aquamarine (Team B matches). Radial bars represent the learned weights for each covariate.}
\label{fig:partidos}
\end{figure}

Figure~\ref{fig:partidos} presents the model's performance during a 10-day prediction window from October 1 to October 10, 2019. The model was trained on six months of data collected from April 1 to October 1, 2019, starting at midnight. This training period was selected due to the frequent occurrence of football matches, which were expected to influence PoS-related tweet activity. As shown, the model aligns closely with the actual tweet volume on the initial days of the prediction period. However, as the window progresses, the predictions tend to converge toward the average observed activity, reflecting a smoothing behavior in the absence of strong external signals.

\noindent Analysis of the fitted covariate weights reveals negative values for the football match indicators, suggesting that, during this period, these events had a limited or even inverse association with PoS-related tweet volume. In contrast, protest events maintained a stronger, positive influence on predicted activity. A notable deviation occurs on the final day of the prediction window, which shows a sharp increase in observed tweets not captured by the model. This surge coincides with the onset of a protest not present in the training data, illustrating a limitation of the predictive model in responding to unforeseen future events.

\subsubsection{Comparison with Other Models}

\begin{figure}[t]
\centering
\includegraphics[width=0.48\textwidth]{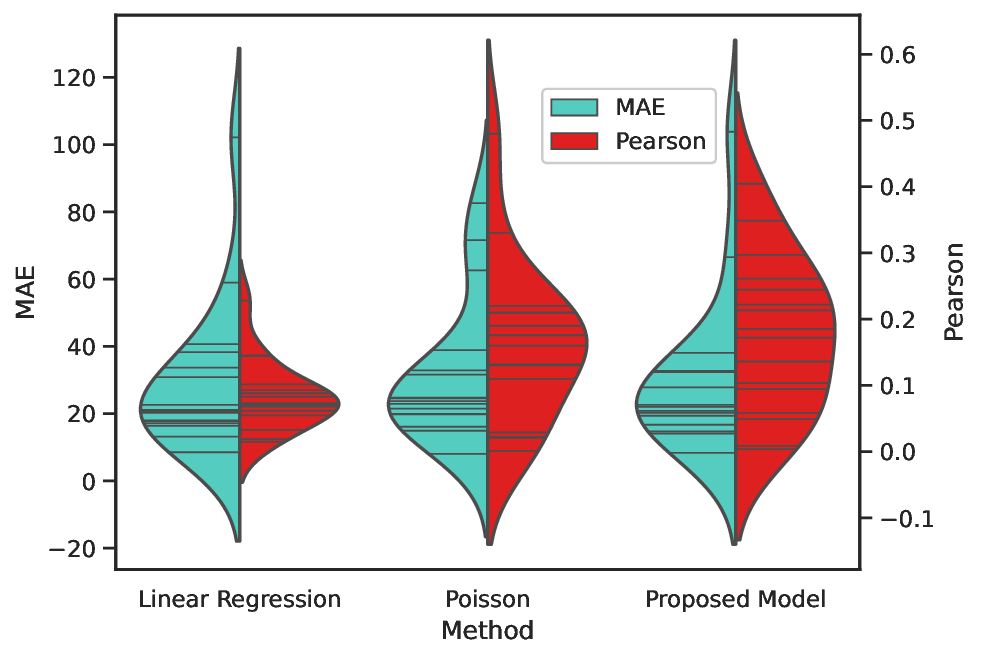}
\caption{Distribution of MAE (left) and Pearson correlation coefficients (right) across cross-validation folds. The training window spans 30 days, with partitions advancing by 15 days. Evaluation is performed over an 8-day prediction window in each fold.}
\label{fig:metrics}
\end{figure}

Figure~\ref{fig:metrics} compares the performance of the proposed Hawkes-based model with two baseline models: a non-homogeneous Poisson process and a linear regression approach. The figure presents violin plots summarizing the distributions of the Mean Absolute Error (MAE) and the Pearson correlation coefficient across multiple cross-validation folds. For the proposed model, the average MAE was $31.01 \pm 23.68$, indicating that, on average, predictions deviated from the observed tweet counts by approximately 31 tweets. The Poisson model yielded an average MAE of $30.48 \pm 21.44$, while the linear regression model produced an average MAE of $30.24 \pm 23.02$. These results suggest comparable predictive performance across all three models, with no statistically significant differences in average error.

\noindent In terms of correlation, the proposed model achieved an average Pearson coefficient of $0.17 \pm 0.12$, indicating a modest positive association between predicted and actual tweet volumes. The Poisson model produced an average of $0.16 \pm 0.13$, while the linear regression model reached $0.08 \pm 0.05$. Although all models show limited correlation strength, the proposed model slightly outperformed the baselines. Importantly, the proposed model provides a distinctive advantage in interpretability. Unlike the baselines, it enables the quantification of covariates' contributions to tweet activity through its parameterization. This feature supports a more informed understanding of the temporal and contextual drivers of PoS-related content on social media. In summary, while the predictive accuracy of all models is comparable, the Hawkes-based model uniquely offers interpretable structure, making it a valuable tool for both prediction and explanatory analysis of social media dynamics.

\section{Discussion}

This work contributes to the growing field of predictive analytics for Perception of Security (PoS) by introducing a temporally-aware, interpretable model capable of forecasting short-term fluctuations in PoS-related discourse on social media. The findings highlight the potential of combining textual sentiment analysis with event-sensitive temporal modeling to inform public decision-making processes.

\subsection{Interpretable Predictive Modeling}

Unlike many black-box approaches, the proposed Hawkes-based model offers a transparent structure in which the influence of contextual and temporal covariates can be explicitly quantified. This feature enables not only short-term forecasting but also post-hoc analysis of the drivers behind variations in public sentiment. In operational contexts, such interpretability may support scenario assessment and proactive planning by public institutions, especially when anticipating responses to recurring events such as civic demonstrations.

\subsection{Influence of Contextual Covariates}

Empirical results showed that protest events were consistently associated with increased PoS-related activity, confirming their salience in shaping online discourse about security. In contrast, the effect of soccer matches was weaker and more variable, underscoring the context-specific nature of their impact. These observations align with prior findings in urban sociology and risk perception, and suggest that different types of events carry asymmetric informational and emotional weight in shaping public narratives about security.

\subsection{Limitations and Future Directions}

While the proposed model performs comparably to baseline methods in terms of prediction accuracy, its main advantage lies in interpretability and contextual adaptability. Nonetheless, several limitations should be noted. First, the data used in this study correspond to a pre-pandemic period (2019–2020), which may limit generalizability to current social dynamics. This constraint reflects the period of data availability in the original predictive analytics project conducted in Bogotá at that time. Second, the model assumes temporal stationarity in covariate effects and does not incorporate mechanisms for handling abrupt, unforeseen events. Future work could integrate adaptive or event-detection components to enhance responsiveness. Third, only two exogenous covariates were considered. While illustrative, this leaves unexplored a wide range of factors (e.g., crime incidents, news media content, weather, policy changes) that may further influence perceived security.

\noindent Finally, the model operates at the aggregate level (tweet volume) and does not explicitly consider user interactions or the network topology of diffusion. Incorporating network-aware features or extending the model to individual-level dynamics could offer deeper insights into the mechanisms through which perceptions are formed and propagated.

\section{Conclusions}

This study proposed an interpretable predictive framework for estimating the volume of tweets related to Perception of Security (PoS) on social media. Built upon a Hawkes point process with contextual covariates, the model not only offers competitive predictive performance when compared to baseline approaches, but also enables the identification and quantification of the influence of external events on online public sentiment. Although the model’s forecasting accuracy, measured through MAE and Pearson correlation, was similar to that of standard methods, its interpretability constitutes a significant advantage. By revealing the role of specific covariates in shaping predicted tweet volumes, the model provides actionable insights for decision-makers and security planners. This work establishes a methodological foundation for future research in predictive analytics applied to public sentiment and urban security. Future directions include enhancing model responsiveness to emerging events, incorporating additional data sources, and extending the framework to different social and geographic contexts. As the field continues to evolve, interpretable models such as the one proposed here offer promising tools for anticipating shifts in public perception and informing timely interventions.

\appendix[Mathematical Derivation and Model Extension]
\label{anexo:modelo_matematico}

This appendix provides the mathematical derivation and formal extension of the proposed model, offering a comprehensive understanding of the equations introduced in the main body of the paper. The objective is to clarify the step-by-step formulation of the model, grounded in a rigorous mathematical foundation. By detailing the structure and logic behind each component, we aim to provide deeper insight into the model's inner workings and the rationale behind its design.

\subsection{Hawkes Point Process}

The core objective of the proposed model is to define the intensity function $\lambda(t)$ over time for the occurrence of security-related posts. Inspired by the Hawkes point process~\cite{rizoiu2017tutorial}, this formulation is well-suited for modeling self-exciting phenomena where events tend to trigger subsequent events. In this context, most tweet-related events are influenced by a combination of two factors: (i) inherent or spontaneous background activity, and (ii) excitations induced by earlier posts. The general form of the Hawkes process is defined as:

\begin{equation}
\label{eq:hawkes_general}
\lambda(t) = \mu(t) + \sum_{i: t_i < t} g(t - t_i).
\end{equation}

\noindent In this formulation, the function $\mu(t)$ represents the background intensity, capturing the inherent temporal dynamics of the platform. Specifically, the spontaneous posting of original tweets. In contrast, the excitation kernel $g(t)$ models the additional intensity contributed by past events, including both original tweets and retweets, which may influence the occurrence of future posts. This decomposition is essential to capture the self-exciting nature of information diffusion and the evolving dynamics of security-related content on the Twitter platform.

\subsubsection{Original Tweet Posting}

The background intensity term $\mu(t)$ in Equation~\ref{eq:hawkes_general} is modeled as an exponential function that dynamically adjusts the baseline rate of original tweet postings based on a set of time-dependent covariates. This formulation is given by:

\begin{equation}
\label{eq:mu_t}
\mu(t) = \exp\left( \boldsymbol{\beta}^\top \mathbf{C}(t) \right).
\end{equation}

\noindent
Here, $\mathbf{C}(t)$ denotes a vector of covariates observed at time $t$, and $\boldsymbol{\beta}$ is the corresponding vector of learned coefficients. The inner product $\boldsymbol{\beta}^\top \mathbf{C}(t)$ determines the logarithm of the background intensity, such that $\mu(t)$ reflects the effect of contextual factors on the expected arrival rate of original tweets. Notably, $\mathbf{C}(t)$ includes an intercept term (i.e., a constant entry equal to one), which enables the model to learn a base intensity independent of the observed covariates.

\subsubsection{Retweet Posting}

In Equation~\ref{eq:hawkes_general}, the summation term represents the cumulative contribution of past original tweets through their retweets. This contribution comprises three key components: (i) the time-varying influence of the original tweet, (ii) the social reach of the retweeting accounts, and (iii) the temporal likelihood of a retweet occurring at time $t$. Mathematically, the excitation effect of an original tweet $i \in OT$ and its associated set of retweets $RT(i)$ is defined as:

\begin{equation}
\label{eq:sum_g}
g(t - t_i) = p_i(t) \sum_{j \in RT(i)} d_j \, \psi(t - t_j).
\end{equation}

\noindent
Here, $p_i(t)$ denotes the instantaneous influence of the original tweet $i$ at time $t$, $d_j$ corresponds to the number of followers of the user responsible for retweet $j$, and $\psi(s)$ is a probability density function modeling the delay distribution between an original tweet and its retweets~\cite{rizoiu2017tutorial,kobayashi2016tideh}.

\noindent Figure~\ref{fig:psi} illustrates the behavior of $\psi(s)$, which remains constant for short lags and then decays over time. Figure~\ref{fig:influence} shows the temporal evolution of $p_i(t)$ for different levels of perceived security (PoS). This influence function is modeled as an exponentially decaying oscillatory function, defined as:

\begin{multline}
\label{eq:influence}
p_i(t) = p_0^i \left[ 1 - (S_i r_0) \sin\left( \frac{2\pi}{T_m}(t + \phi_0) \right) \right] \\
\times \exp\left( -\frac{t - t_0}{\tau_m} \right).
\end{multline}

\noindent The objective is to estimate the parameter set $\Theta = \{P_0, r_0, \phi_0, \tau_m, \boldsymbol{\beta}\}$ from historical data. To this end, we employ a maximum likelihood estimation (MLE) framework following the approach described in~\cite{reinhart2016point}.

\subsection{Estimating Model Parameters}

Following the principles outlined in~\cite{reinhart2016point}, for a Hawkes point process as defined in Equation~\ref{eq:hawkes_general}, the maximum likelihood estimator (MLE) of the parameter set $\Theta$ over a set of $n$ historical events occurring within the time interval $[t_a, t_b]$ is given by:

\begin{equation}
L(\Theta) = \left( \prod_{i=1}^{n} \lambda(t_i) \right) \exp\left( - \int_{t_a}^{t_b} \lambda(t) \, \mathrm{d}t \right).
\end{equation}

\noindent Let $\ell(\Theta) = \ln(L(\Theta))$ be the log-likelihood. Applying the properties of the logarithmic function, the expression becomes:

\begin{equation}
\label{eq:loglike}
\ell(\Theta) = \sum_{i=1}^{n} \ln\left( \lambda(t_i) \right) - \int_{t_a}^{t_b} \lambda(t) \, \mathrm{d}t.
\end{equation}

\noindent In the context of the proposed model, the original tweets and their corresponding retweets are explicitly observed. According to the modeling assumption, the background intensity $\mu(t)$ governs the generation of original tweets, whereas the excitation component $g(t)$ accounts for the influence of retweets. To separate these components in the likelihood expression, we define an auxiliary function $u_i$ as follows:

\begin{equation}
\label{eq:aux_eq}
u_i =
\begin{cases}
0 & \text{if } i \in OT \\
j & \text{if } j \in RT(i).
\end{cases}
\end{equation}

\noindent
By applying the indicator function $\mathbbm{1}$ and using Equation~\ref{eq:aux_eq} in the log-likelihood expression from Equation~\ref{eq:loglike}, we obtain:

\begin{equation}
\label{eq:loglike2}
\begin{split}
\ell(\Theta) = &\sum_{i=1}^{n} \mathbbm{1}(u_i = 0)\, \ln\left( \mu(t_i) \right) \\
&+ \sum_{i=1}^{n} \sum_{j=1}^{n} \mathbbm{1}(u_i = j)\, \ln\left( g(t_i - t_j) \right) \\
&- \int_{t_a}^{t_b} \lambda(t)\, \mathrm{d}t.
\end{split}
\end{equation}

\noindent
In Equation~\ref{eq:loglike2}, the first summation captures the contribution of original tweets ($OT$), while the second summation accounts for retweet events. By systematically associating these components and substituting the expressions from Equations~\ref{eq:sum_g} and~\ref{eq:influence}, we obtain the final expression for the log-likelihood:

\begin{equation}
\label{eq:loglike3}
\begin{split}
\ell(\Theta) = &\sum_{i \in OT} \boldsymbol{\beta}^\top \mathbf{C}(t_i) \\
&+ \sum_{i \in OT} \sum_{j \in RT(i)} \ln\left( p_i(t_j)\, d_j\, \psi(t_j - t_i) \right) \\
&- \int_{t_a}^{t_b} \lambda(t)\, \mathrm{d}t.
\end{split}
\end{equation}

\subsubsection{Estimating Background Parameters}

It is important to note that the parameter vector $\boldsymbol{\beta}$ only affects the first term in Equation~\ref{eq:loglike3}, as well as the first component of the integral term when substituting $\lambda(t)$. By computing the gradient of the log-likelihood $\ell(\Theta)$ with respect to $\boldsymbol{\beta}$ and setting it equal to zero, we obtain:

\begin{equation}
\label{eq:beta}
\frac{\partial \ell}{\partial \boldsymbol{\beta}} = \sum_{i \in OT} \mathbf{C}(t_i)
- \frac{\partial \ell}{\partial \boldsymbol{\beta}} \left( \int_{t_a}^{t_b} \exp\left( \boldsymbol{\beta}^\top \mathbf{C}(t) \right)\, \mathrm{d}t \right) = 0.
\end{equation}

\noindent Given that the covariate function $\mathbf{C}(t)$ is typically piecewise constant, e.g., changing only by hour or day, it is possible to define a partition $\mathcal{P}$ of the interval $[t_a, t_b]$ such that $\mathbf{C}(t)$ remains constant over each subinterval $p_i \in \mathcal{P}$. Denoting by $\overline{p}_i$ the midpoint of $p_i$, Equation~\ref{eq:beta} can be rewritten as:

\begin{equation}
\label{eq:beta2}
\begin{split}
\sum_{i \in OT} \mathbf{C}(t_i) &= \sum_{p_i \in \mathcal{P}} \exp\left( \boldsymbol{\beta}^\top \mathbf{C}(\overline{p}_i) \right)\, |p_i|\, \mathbf{C}(\overline{p}_i).
\end{split}
\end{equation}

\noindent Here, $|p_i|$ denotes the length of the partition interval $p_i$, and $\overline{p_i}$ represents its midpoint. From this expression, a closed-form solution for $\boldsymbol{\beta}$ is not readily available. Therefore, numerical methods are employed to approximate its value. It is worth noting that the original likelihood function is convex with respect to $\boldsymbol{\beta}$~\cite{reinhart2016point}, which facilitates the optimization process and ensures convergence to a global optimum.

\subsubsection{Estimating Influence Function Parameters}

The remaining parameters in $\Theta$ are estimated using the maximum likelihood method; however, the estimation is not performed directly from Equation~\ref{eq:influence}. Instead, we assume that the influence function $p_i(t)$ remains constant within short time windows, and approximate it by a discrete version denoted as $\widehat{p}_i(t)$. The estimation procedure applies MLE to $\widehat{p}_i(t)$ and minimizes the discrepancy between the analytical form $p_i(t)$ and its empirical counterpart $\widehat{p}_i(t)$. Specifically, for each $i \in OT$, within a given time window $\Delta t = t_{\mathrm{end}} - t_{\mathrm{st}}$ in which $\widehat{p}_i(t)$ is assumed constant, the log-likelihood in Equation~\ref{eq:loglike3} becomes:

\begin{equation}
\label{eq:loglike4}
\begin{split}
\ell(\Theta; \Delta t) &= \sum_{\substack{i \in OT \\ i \in \Delta t}} \boldsymbol{\beta}^\top \mathbf{C}(t_i) \\
&\quad + \sum_{\substack{i \in OT \\ i \in \Delta t}} \sum_{j \in RT(i)} \ln\left( \widehat{p}_i(t_j)\, d_j\, \psi(t_j - t_i) \right) \\
&\quad - \int_{t_{\mathrm{st}}}^{t_{\mathrm{end}}} \lambda(t)\, \mathrm{d}t.
\end{split}
\end{equation}

\noindent The derivative of $\ell(\Theta; \Delta t)$ with respect to $\widehat{p}_i$ is given by:

\begin{equation}
\label{eq:derivate_pi}
\frac{\partial \ell}{\partial \widehat{p}_i} =
\sum_{\substack{j \in RT(i) \\ j \in \Delta t}} \frac{1}{\widehat{p}_i}
- \int_{t_{\mathrm{st}}}^{t_{\mathrm{end}}}
\sum_{\substack{j \in RT(i) \\ j \in \Delta t}} \widehat{p}_i\, d_j\, \psi(t - t_j)\, \mathrm{d}t.
\end{equation}

\noindent Setting Equation~\ref{eq:derivate_pi} equal to zero and applying the change of variables $x = t - t_j$, we obtain:

\begin{equation}
\label{eq:p_i_fit}
\widehat{p}_i (\Delta t) = R_i \left( \sum_{\substack{j \in RT(i) \\ j \in \Delta t}} d_j \int_{t_{\mathrm{st}} - t_j}^{t_{\mathrm{end}} - t_j} \psi(t)\, \mathrm{d}t \right)^{-1}.
\end{equation}

\noindent
where $R_i$ is the number of retweets of tweet $i$ that fall within the time window $\Delta t$. Now, for each $i \in OT$, let $D(i)$ denote the sequence of times $t_i = d_0 < d_1 < \ldots < d_{k(i)}$, where $d_{k(i)}$ is the timestamp of the last retweet in $RT(i)$. These points define the partition into subintervals $\Delta t_1, \ldots, \Delta t_{k(i)}$, over which $\widehat{p}_i$ is assumed constant.

\noindent The fitting error function $E(P_0, r_0, \phi_0, \tau_m)$ is then defined as:

\begin{equation}
E(P_0, r_0, \phi_0, \tau_m) =
\sum_{i \in OT} \sum_{j = 1}^{k(i)} \left\| \widehat{p}_i(\Delta t_j) - p_i(\Delta t_j) \right\|.
\end{equation}

\noindent
where $p_i(\Delta t_j)$ is evaluated at the midpoint of the interval $\Delta t_j$. This error function is minimized to estimate the parameters $P_0$, $r_0$, $\phi_0$, and $\tau_m$.

\subsubsection{Predicting Future Tweets}

Once the model parameters have been estimated, they can be used to predict future tweet activity, including both original tweets and their corresponding retweets, over a future interval $[t_b, t_p]$. However, in Equation~\ref{eq:influence}, the values $p_0^i$, $S$, and $t_0$ are only known for tweets $i \in OT$ observed during the training period. Similarly, the number of followers $d_j$ is only available for retweets within $[t_a, t_b]$. To address this limitation, we proceed as follows:

\begin{itemize}
    \item Using Equation~\ref{eq:mu_t} and the estimated coefficients $\boldsymbol{\beta}$, original tweets (denoted as $OTP$) are sampled over the interval $t \in [t_b, t_p]$ using the thinning algorithm~\cite{rizoiu2017tutorial}.
    
    \item For each sampled original tweet $i \in OTP$, the posting time $t_0$ is recorded. Then, values for $p_0^i$ and $S$ are drawn from the empirical distributions of these variables in the training data. This allows the computation of $p_i(t)$ for any $t \in [t_b, t_p]$ using Equation~\ref{eq:influence}.
    
    \item Although the original tweets are simulated, the behavior of their retweets is unknown. Therefore, the expected intensity contributed by retweets is approximated using the expected number of followers $\mathbb{E}(d_i)$ and the cumulative value of the response kernel $\psi(t)$.
\end{itemize}

\noindent
Finally, the influence of previously observed events (from the training period) is still evaluated via $\lambda(t)$. Thus, for any $t > t_b$, the predicted total intensity is given by:

\begin{multline}
\label{eq:lambda_future}
\mathbb{E}[\lambda(t) \mid OTP] = \widehat{\lambda}(t) = \lambda(t) \\
+ \sum_{i \in OTP} \mathbb{E}(p_i(t))\, \mathbb{E}(d_i) \int_{t_b}^{t} \psi(t - s)\, \mathrm{d}s.
\end{multline}

\noindent
Once the expected intensity $\widehat{\lambda}(t)$ has been computed, the thinning algorithm~\cite{rizoiu2017tutorial} is applied again to generate predicted tweet events within the interval $[t_b, t]$.

\section*{Acknowledgment}

This work was funded by the project Dise{\~n}o y validaci{\'o}n de modelos de anal{\'i}tica predictiva de fen{\'o}menos de seguridad y convivencia para la toma de decisiones en Bogot{\'a} (BPIN: 2016000100036).

\noindent The authors also acknowledge the use of OpenAI's ChatGPT for language editing and writing assistance. The system was used to improve clarity, structure, and academic style in several sections of the manuscript, including the Introduction, Abstract, and Discussion. All AI-generated suggestions were critically reviewed and validated by the authors prior to inclusion.

\section*{Conflict of Interest}

The authors declare that they have no conflict of interest regarding the publication of this paper.


\bibliographystyle{IEEEtran}
\bibliography{bibtex/bib/mybibfile}
%

%








\end{document}